\documentclass[aps,prc,groupedaddress,reprint,showpacs,floatfix]{revtex4-1}
\usepackage{graphicx}
\begin{document}


\title{Measurement of alpha induced reaction cross-sections on $^{nat}$Mo with detailed covariance analysis}


\author{Mahesh Choudhary}\email{maheshchoudhary921@gmail.com}
\author{A. Gandhi}
\author{Aman Sharma}
 \author{Namrata Singh}
\author{A. Kumar}\email{ajaytyagi@bhu.ac.in}
\affiliation{Department of Physics, Banaras Hindu University, Varanasi-221005, India}
\author{S. Dasgupta} 
\author{J. Datta}
\affiliation{ Analytical Chemistry Division, BARC, Variable Energy Cyclotron Center, Kolkata-700064, India }
\date{\today}
\begin{abstract}
In the present study we have measured the excitation functions for the nuclear reactions  $^{100}$Mo($\alpha$,n)$^{103}$Ru, $^{nat}$Mo($\alpha$,x)$^{97}$Ru, $^{nat}$Mo($\alpha$,x)$^{95}$Ru,  $^{nat}$Mo($\alpha$,x)$^{96g}$Tc, $^{nat}$Mo($\alpha$,x)$^{95g}$Tc and $^{nat}$Mo($\alpha$,x)$^{94g}$Tc in the energy range 11-32 MeV. We have used the stacked foil activation technique followed by off-line gamma ray spectroscopy technique to measure the excitation functions. In this study we have also documented detailed uncertainty analysis for these nuclear reactions and their corresponding covariance matrix are also presented. The excitation functions are compared with the available experimental data from EXFOR data library  and the theoretical prediction from TALYS nuclear reaction code. The present measurements are found to be consistent with the available experimental data.  
\end{abstract}
\maketitle

\section{Introduction}

The cross-sections measurement for the  charged particle-induced reactions with various target materials are of great interest in many fields such as astrophysics, production of medical radioisotopes, nuclear reaction studies, investigations and improvement of physical properties \cite{01,02}. Nowadays, there are various radioactive materials with unique properties, playing an important role in nuclear medicine diagnosis \cite{03, 04}. Out of such radioisotopes, $^{97}$Ru $(T_{1/2}$=2.79 days) is the one that emits low energy (E$_\gamma$ = 215.70 keV) $\gamma$-ray with a branching ratio (I$_\gamma$) $85.2\%$ that makes it suitable for use in single photon emission computed tomography (SPECT) imaging \cite{05}.
The SPECT is a nuclear medicine tomographic imaging technique using $\gamma$-rays that is capable of providing true 3-D image.
An another important radioisotope is $^{103m}$Rh with short half-life of 56.11 min, which is useful for augur electron therapy due to it’s very low photon/electron ratio \cite{06}. The radioisotope $^{103}$Ru has a half-life of 39.26 days which decays via $\beta^-$ decay into $^{103m}$Rh with a branching ratio (I$_\gamma$) of $91\%$. Since $^{103m}$Rh has a short half-life of 56.11 min, $^{103}$Ru as a parent to $^{103m}$Rh provides a good option due to its relatively longer half-life of 39.26 days.
Alpha-induced reactions with molybdenum target are important for the production of the above-mentioned radioisotopes and these nuclear reactions also play an important role in astrophysics for the study of p-process and r-process nucleosynthesis. \cite{07}. The nuclear reaction $^{100}$Mo($\alpha$,n)$^{103}$Ru is used to study the weak r-process nucleosynthesis \cite{08}. In the present study, we have used alpha particles as the projectile and molybdenum as the target material. The main goal of this investigation was to extend the activation cross-section data for nuclear reactions $^{nat}$Mo($\alpha$,x)$^{97}$Ru, $^{100}$Mo($\alpha$,n)$^{103}$Ru  with respect to the production of medicines related to $^{97}$Ru, $^{103}$Ru.\\
Since, with the increasing demand for alpha-induced reactions in nuclear research, astrophysics, nuclear medicine diagnostics, industry, it is an important component to accurately estimate the amount of uncertainty propogation from the evaluated nuclear reaction cross section data of these reactions. Although experimental data for these reactions exist in the EXFOR data library, yet a detailed covariance analysis is not available with any of the data. Therefore, it is important to discuss covariance analysis in detail. 
Covariance analysis is a method used to estimate the uncertainty in the measured quantity with considering cross-correlations, the quantities measured in this study are nuclear reaction cross-sections \cite{09, 10, 11, 12}. In the present work, we provide covariance analysis of nuclear reactions $^{100}$Mo($\alpha$,n)$^{103}$Ru, $^{nat}$Mo($\alpha$,x)$^{97}$Ru, $^{nat}$Mo($\alpha$,x)$^{95}$Ru, $^{nat}$Mo($\alpha$,x)$^{96g}$Tc, $^{nat}$Mo($\alpha$,x)$^{95g}$Tc and $^{nat}$Mo($\alpha$,x)$^{94g}$Tc, considering micro-correlations between measurements such as $\gamma$-ray abundance, $\gamma$-ray count, incident flux, detector efficiency  etc.
For the first time covariance analysis has been performed for the present nuclear reaction reactions. We measured the reaction cross-sections for the above mentioned reactions in the incident energy range 11–32 MeV, for this we have used the stacked foil activation technique. In the stacked foil activation technique, a stack of several thin foils is irradiated by an integrated-particle beam along with the monitor foil. After irradiation, the irradiated foils are placed in front of the detector for measurement of the decay gamma from the radionuclide.
The excitation functions are compared with existing experimental data available in the EXFOR  data library and with theoretical calculations performed using the TALYS  nuclear reaction code \cite{13, 14, 15, 16}. The TALYS nuclear reaction code is a computer code system for nuclear reaction calculations based on the Hauser–Fesbach statistical model. This code has separate options for level density and optical model parameters \cite{17, 18}.\\
The present study is organized into the following six sections. The details about experimental technique and experimental setup in section II, the details about efficiency calibration of the detector in section III, the details about covariance analysis, estimation of nuclear reaction cross-section and theoretical calculation in section IV, results and discussion in section V, the conclusion in section VI are presented. 
\section{Experimental technique and Details} 
The experiment was performed at the K-130 cyclotron, Variable Energy Cyclotron Center (VECC), Kolkata, India. The alpha particles was generated from ionized helium. In this experiment, alpha particles were accelerated up to 32 $\pm$ 0.20 MeV and the average beam current was 170 nA.\\
\begin{figure}[b]
\begin{center}
\includegraphics[width=7.5 cm]{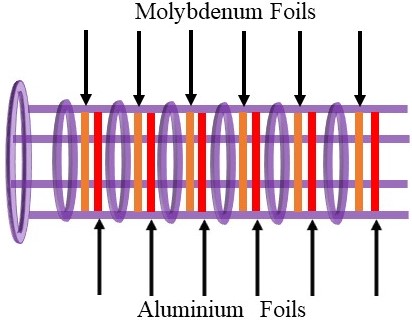}
\caption{Detailed structure of the stacked foil arrangement.}
\end{center}
\end{figure}
The stacked foil activation \cite{19, 20, 21, 22} technique has been used to measure the reaction cross-section of alpha-induced reactions on molybdenum in the energy range from the threshold energy of contributing reaction up to 32 MeV. In this method, a series of thin target foils are placed to form the target as shown in Fig. 1 and each target foil ($^{nat}$Mo in the present work) is followed by another material ($^{nat}$Al in the present work) to catch  the recoils nuclides from the target foil ($^{nat}$Mo). This catcher foil must be such that is does not produce any radioactive product in the incident particle energy range and it has also as low Z-material as possible to decrease the gamma attenuation during the measurements. Therefore a pair of Mo an Al foils will contain the total produced radioactive isotopes from the target Mo foil after irradiation. A detailed structure of the stacked foil arrangement is shown in Fig. 1. The main advantage of the stacked foil activation method is that we can find the complete excitation function curve and each target foils are irradiated with the from a single beam. The Mo and Al foils were measured simultaneously to capture the recoil products and to calculate the nuclear reaction cross sections of the products formed in the Mo foils. The energy drop in the stacked foils were calculated using the Stopping and Range of Ions in Matter (SRIM) code \cite{23, 24}.  The SRIM is a code that provides us details about the energy drop of incident ions and the range in the matter.
Three thin metallic foils of $^{nat}$Mo, $^{nat}$Al and $^{nat}$Ti of thickness 12.85 $mg/cm^2$ , 13.5 $mg/cm^2$ and 1.80 $mg/cm^2$ respectively were used. We made a set of stacks by placing molybdenum foils followed by aluminum foils as the Mo-Al (10 x 10 $mm^2$), and have been used  6 such sets to make a stacked target. In this experiment, the Ti foil (10 x 10 $mm^2$ ) was used as the monitor foil to confirm the incident beam intensity and energy.
The stacked target was irradiated for 2 hours by a 32 MeV alpha beam. Two collimators of 7 mm diameter were used in this experiment, one placed in front of the beam and the other placed in front of the stacked target. At the end of the experiment setup, a Faraday cup was attached for current measurement. A schematic diagram of the irradiation setup is shown in Fig.  2.
\begin{figure}[t]
\begin{center}
\includegraphics[width= 9 cm]{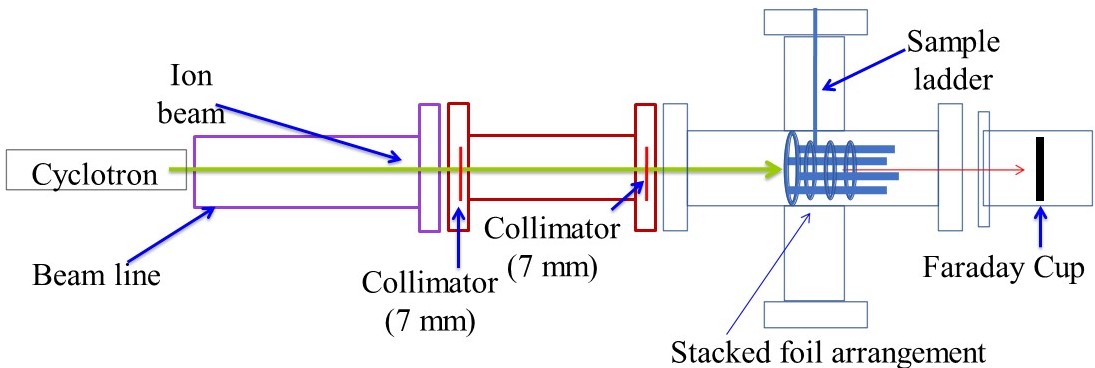}
\caption{The Schematic diagram of experimental setup.}
\end{center}
\end{figure}
\begin{table*}[t]
\begin{center}
{\ TABLE I. The value of the fitting parameter of efficiency ($\varepsilon$) curve and their uncertainty\\
\vspace{2mm}

\begin{tabular}{ccccccc}
\hline
\hline
Parameters&~~~~~~~~~~~~~~~~~~~~~Value&~~~~~~~~~~~~~~~~~~~~~Uncertainty\\
\hline
\vspace{1mm}
$\varepsilon_{c}$ & ~~~~~~~~~~~~~~~~~~~~~0.00137 & ~~~~~~~~~~~~~~~~~~~~~6.2907$\times10^{-5}$  \\
\vspace{1mm}
$\varepsilon_{0}$ & ~~~~~~~~~~~~~~~~~~~~~0.00912 & ~~~~~~~~~~~~~~~~~~~~~4.2340$\times10^{-4}$\\
\vspace{1mm}
E$_{0}$ (keV) & ~~~~~~~~~~~~~~~~~~330.9628 & ~~~~~~~~~~~~~~~~~~20.4224\\
\hline
\hline
\end{tabular}}
\end{center}
\end{table*}
\begin{table*}[t]
\begin{center}
{\ TABLE II. Coincidence summing effect, correction factor, and efficiency of HPGe detectors at different gamma-ray energies with their intensities for both sample ($\varepsilon$) and point source ($\varepsilon_{p}$) geometries.}
\vspace{2mm}

\begin{tabular}{cccccc}
\hline
\hline
$E_{\gamma}$(keV)&~~~~~~~~~~$I_{\gamma}$&~~~~~~~~~~Counts(C)&~~~~~~~~~~K$_{c}$&~~~~~~~~~~$\varepsilon_{p}$ &~~~~~~~~~~$\varepsilon$ \\
\hline
\vspace{1mm}
121.78 &~~~~~~~~~~0.2853 $\pm$ 0.0016&~~~~~~~~~~ 509811 $\pm$ 764.71 &~~~~~~~~~~1.032& ~~~~~~~~~~0.027099 &~~~~~~~~~~0.007785 $\pm$ 0.0000605 \\
\vspace{1mm}
244.69 &~~~~~~~~~~0.0755 $\pm$ 0.0004 &~~~~~~~~~~ 93725 $\pm$ 337.41 &~~~~~~~~~~1.041 &~~~~~~~~~~0.018990 &~~~~~~~~~~0.005726 $\pm$ 0.0000471 \\
\vspace{1mm}
344.27 &~~~~~~~~~~0.2659 $\pm$ 0.0020 &~~~~~~~~~~ 265271 $\pm$ 530.54 &~~~~~~~~~~1.022 &~~~~~~~~~~0.014982 &~~~~~~~~~~0.004589 $\pm$ 0.0000429 \\
\vspace{1mm}
411.11 &~~~~~~~~~~0.02238 $\pm$ 0.00013 &~~~~~~~~~~17985 $\pm$ 160.06 &~~~~~~~~~~1.050 &~~~~~~~~~~0.012399 &~~~~~~~~~~0.003820 $\pm$ 0.0000451 \\
\vspace{1mm}
688.70 &~~~~~~~~~~0.0085 $\pm$ 0.0008 &~~~~~~~~~~4415 $\pm$ 94.48 &~~~~~~~~~~1.007 &~~~~~~~~~~0.007686 &~~~~~~~~~~0.002401 $\pm$ 0.0002321 \\
\vspace{1mm}
867.38 &~~~~~~~~~~0.0423 $\pm$ 0.0003 &~~~~~~~~~~18156 $\pm$ 152.51 &~~~~~~~~~~1.045 &~~~~~~~~~~0.006591 &~~~~~~~~~~0.002070 $\pm$ 0.0000252\\
\vspace{1mm}
964.05 &~~~~~~~~~~0.1451 $\pm$ 0.0007 &~~~~~~~~~~62252 $\pm$ 261.45 &~~~~~~~~~~1.017 &~~~~~~~~~~0.006411 &~~~~~~~~~~0.002019 $\pm$ 0.0000166 \\
\vspace{1mm}
1085.83 &~~~~~~~~~~0.1011 $\pm$ 0.0005 &~~~~~~~~~~36755 $\pm$ 205.82 &~~~~~~~~~~0.986 &~~~~~~~~~~0.005267 &~~~~~~~~~~0.001663 $\pm$ 0.0000151 \\
\vspace{1mm}
1112.94 &~~~~~~~~~~0.1367 $\pm$ 0.0008 &~~~~~~~~~~49937 $\pm$ 236.20 &~~~~~~~~~~1.007 &~~~~~~~~~~0.005405 &~~~~~~~~~~0.001708 $\pm$ 0.0000156 \\
\vspace{1mm}
1299.1 &~~~~~~~~~~0.0163 $\pm$ 0.00011 &~~~~~~~~~~5351 $\pm$ 80.26 &~~~~~~~~~~1.030 &~~~~~~~~~~0.004968 &~~~~~~~~~~0.001575 $\pm$ 0.0000272 \\
\vspace{1mm}
1408.01 &~~~~~~~~~~0.2087 $\pm$ 0.0009 &~~~~~~~~~~64230 $\pm$ 256.92 &~~~~~~~~~~1.012 &~~~~~~~~~~0.004576 &~~~~~~~~~~0.001453 $\pm$ 0.0000114 \\
\hline
\hline
\end{tabular}
\end{center}
\end{table*}
\section{Gamma-ray spectrometry}
The activity of the samples were measured using high purity germanium detector (HPGe) having 40 $\%$ relative efficiency. In the present work, the standard $^{152}Eu$ point source has been used to calculate the efficiency of the HPGe detector for various gamma-ray energies. The half-life of the standard $^{152}Eu$ point source is T$_{1/2}$ = 13.517 $\pm$  0.009 years and the known activity was reported on 17 May 1982 which is A$_0 = 3.908 \times10^{4}$ Bq.
\begin{figure}[b]
\begin{center}
\includegraphics[width=8.8 cm]{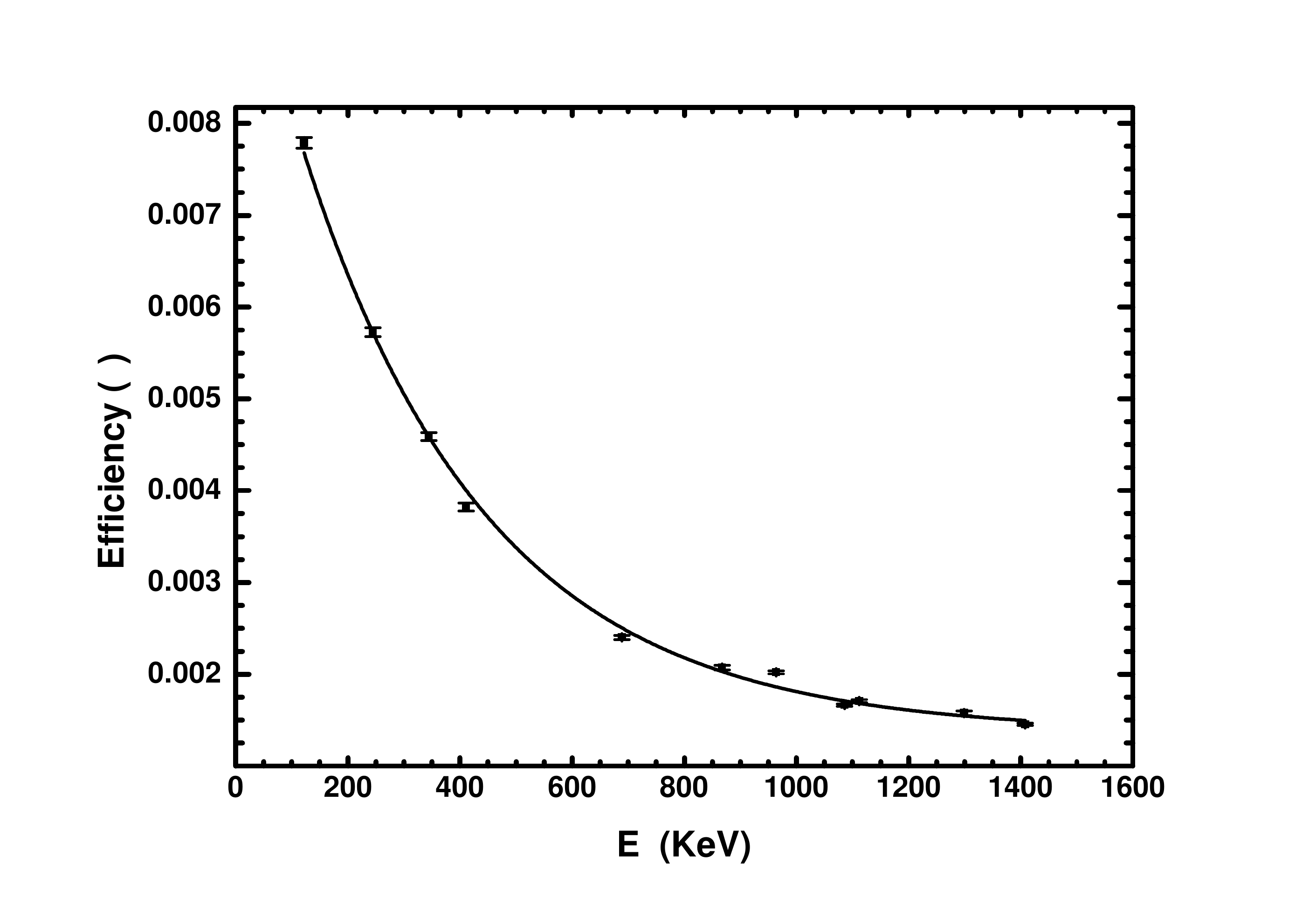}
\caption{Efficiency curve of HPGE detector for a distance of 50 mm between source and detector absorber.}
\end{center}
\end{figure}
The absolute efficiency calibration of the point source for the source-detector distance 50 mm was estimated using the following equation \cite{25}: 
\begin{equation}
\varepsilon_{p} = \frac{CK_{c}}{A_{0}I_{\gamma}\Delta{t}e^{-{\lambda}t}} 
\end{equation}
in above equation, A$_{0}$ is the known activity of standard $^{152}Eu$ point source, the total number of count C was taken in 10000  second for ${\gamma}$-ray energy with absolute intensity ($I_\gamma$) and t is the time interval from the date of a manufacturing point source to the start of measurement. Here, the summing correction factor is $K_C$ . When the distance between the source and absorber of the detector was 50 mm, shows a coincidence summing effect, so it should be corrected for efficiency calibration of the detector. The standard sample $^{152}Eu$ was a point source and our samples have a finite area, so the efficiency of the point source geometry ($\epsilon_p$) has to be changed for the efficiency of the sample geometry ($\epsilon$). We have used the Monte Carlo simulation code EFFTRAN \cite{26, 27} to calculate the correction factor ($K_C$) of coincidence summing effect and convert the efficiency of the point source ($\epsilon_p$) geometry to the efficiency of the sample ($\epsilon$) geometry.\\
The calculated efficiency value for sample source geometry ($\epsilon$) and point source geometry ($\epsilon_p$) placed at 50 mm from the detector absorber are given in Table II with the correction factor (K$_C$) of coincidence summing effect. 
To calculate the efficiency of a particular $\gamma$-ray of the product radionuclide, we have used equation 2 which is a fitting function of interpolating the point-wise efficiencies ($E_{\gamma}$) of the $\gamma$-ray energy of the standard source $^{152}Eu$ \cite{25, 28}.
\begin{equation}
\varepsilon(E_{\gamma}) = \varepsilon_{o}exp({-E_{\gamma}/E_{0}}) + \varepsilon_{c}
\end{equation}
In the above equation  $\varepsilon_{o}$, $\varepsilon_{c}$, and $E_{0}$ are the fitting parameters of the measured detection efficiencies ($\varepsilon$) of standard point $^{152}Eu$. The value of the curve fitting parameters mentioned in equation 2 are given in Table I. 
\section{DATA ANALYSIS}
\begin{center}
\bf{A. Estimation of the reaction cross section}
\end{center}
In this work, the nuclear reaction cross-sections were calculated using the following standard activation formula;

\begin{equation}
\sigma=\frac{C_{\gamma}\lambda}{\varepsilon(E_\gamma){I_\gamma}D_t{\phi}N_te^{-\lambda t_c}(1-e^{-\lambda t_{irr}})(1-e^{-\lambda t_{m}})}
\end{equation}

In the above formula \cite{29}, the peak area counts is $C_\gamma$ of a particular $\gamma$-ray with it’s abundance ($I_{\gamma}$), $\varepsilon(E_\gamma)$ is the detector efficiency with dead time $D_t$, {$\lambda$} is the decay constant ($s^{-1}$), $N_t$ (($cm^{-2}$)) is the particle density in the target material,  {$\phi$} is the incident particle flux per unit time($s^{-1}$).
In equation 3, the cooling time of the irradiated target is $t_c$(s) with the irradiation time $t_{irr}$(s) and the measurement time $t_m$(s) of the sample source.
In the present study, the uncertainty in the factors contributing to the cross-section such as the efficiency of the detector $\varepsilon(E_\gamma)$, gamma-ray intensity ($I_{\gamma}$), number density of atoms ($N_t$), gamma-ray peak counts ($C_\gamma$) etc. is also taken into account to calculate the uncertainty of the obtained cross-section.
\begin{center}
\bf{B. Covariance Analysis }
\end{center}
A covariance analysis is a mathematical tool that can help to describe the detailed uncertainty with the cross-correlation between different measured quantities. The principle of covariance emphasizes the formulation 
of  physical laws using only measurements of physical quantities that can be correlated. 
The covariance matrix of cross-sections I$_\sigma$ can be written as \cite{30, 31};
\begin{equation}
I_\sigma = H_xC_xH_x^T
\end{equation}
Where  I$_\sigma$ is the covariance matrix of measured reaction cross-sections with order m $\times$ m, $C_x$ is the covariance matrix with order n $\times$ n of different attributes  in activation formula (equation 2) like as $\gamma$-ray intensity ($I_\gamma$), detector efficiency $\varepsilon(E_\gamma)$, incident flux ($\phi$), counts of the peak area ($C_\gamma$), decay constant ($\lambda$) etc.
In equation (4), $H_x$ is the sensitivity matrix with element;
\begin{equation}
H_{xij} = \frac{\partial \sigma_i}{\partial x_j}; (i = 1,2,3,...m;~j = 1,2,3,...n) 
\end{equation}
 Here m has value equal to total number of measured cross-sections for a particular nuclear reaction and has value equal to total number of attributes in the activation formula.
 Let us consider $x_j$ (j = 1,2,3....n), various variable parameters (which are mentioned above) required in the calculation of the cross-sections, then the covariance matrix (c$_x$) of these variables can be obtained from the following equation \cite{32};
 \begin{equation}
C_x(x_j,x_k) = Cor(x_j,x_k)(\Delta x_j\Delta x_k) 
\end{equation}
Where $Cor(x_j,x_k)$ is the correlation coefficient between two variable $x_j$, $x_k$ and it has value $0\leq Cor(x_j,x_k) \leq 1$. $Cor(x_j,x_k)$ = 1, when j = k i.e., these two variable $x_j$, $x_k$ are fully correlated. 
\begin{center}
\bf{C. Theoretical Calculations}
\end{center}
The theoretical results for the reactions $^{100}$Mo($\alpha$,n)$^{103}$Ru, $^{nat}$Mo($\alpha$,x)$^{97}$Ru, $^{nat}$Mo($\alpha$,x)$^{95}$Ru,  $^{nat}$Mo($\alpha$,x)$^{96g}$Tc, $^{nat}$Mo($\alpha$,x)$^{95g}$Tc and $^{nat}$Mo($\alpha$,x)$^{94g}$Tc has been done using the statistical nuclear model code TALYS. The TALYS code is a Fortran based theoretical nuclear reaction model code and used for calculating various physical observable related to nuclear reactions.  In the TALYS nuclear code, we can simulate the nuclear reactions that involve neutrons, photons, protons, deuterons, tritons, $^3$He- and alpha-particles as a projectile, in the 1 keV - 200 MeV energy range and for target nuclides of mass 12 and heavier.
The main purpose of TALYS nuclear code is to provide a complete set of answers for the nuclear reactions, all open channels and associated cross sections, spectra and angular distributions. In the TALYS nuclear code, there are 6 level densities models in which three options for the phenomenological level density model and three options for the microscopic level density \cite{33, 34}. Of these six level density models, ldmodel-1 is the constant temperature and Fermi gas model, ldmodel-2 is the back-shifted Fermi gas model, ldmodel-3 is the generalized superfluid model, ldmodel-4 is the microscopic level density (Skyrme Force) from the Gorley table, ldmodel-5 is the microscopic level density (Skyrem force) from Hilaire’s combinatorial tables and ldmodel-6 is the microscale density (temperature-dependent HFB, Gogni force) from Hilaire’s combinatorial tables \cite{35, 36, 37, 38, 39, 40}. The theoretical calculations for present nuclear reaction cross sections, we have used all these six level density models and also presented a comparison of the result of theoretical calculations using these different level density models with experimental data. 
\section{RESULTS AND DISCUSSION}
In this section, we have presented measured reaction cross-sections with their uncertainties and covariance matrix of  $^{nat}$Mo($\alpha$,x) reactions in the energy range from the respective threshold for each contributing reaction to about 32 MeV. The obtained reaction cross-section in the present work are plotted in Fig. 4-9 along with the available experimental data from the EXFOR and theoretical results from the TALYS-1.9 code.The details of the reactions and decay data of the measured radioisotopes are shown in Table III and the value of the measured cross-sections for the present reactions along with their covariance matrix are presented in Tables IV-IX. Whichever level density model gives the best theoretical result along with the experimental data, the theoretical result is shown in a solid line, and other theoretical results are shown in a dash-dash line with different colors in the graph. The experimental data are shown in dark black circles and available experimental data in EXFOR are shown in different shapes with their error bar along the y-axis.

\begin{table*}
\begin{center}
{\ TABLE III. Reactios and the decay data of the measured radioisotopes.}\\
\vspace{2mm} 

\begin{tabular}{ccccccc}
\hline
\hline
Radionuclide&~~~~~~~~~~~~Half-life&~~~~~~~~~~~~$E_{\gamma }$&~~~~~~~~~~~~$I_{\gamma }$~~~~~~~~~~~~&Reaction&~~~~~~~~~~~~Q-value\\
\vspace{2mm}
&~~~~~~~~~~~~($t_{1/2}$)&~~~~~~~~~~~~(keV)&(\%)&&~~~~~~~~~~~~(MeV)\\
\hline
\vspace{1mm}
$^{95}$Ru &~~~~~~~~~~~~1.643 h &~~~~~~~~~~~~336.40 & 69.90 & $^{92}$Mo($\alpha$,n) & ~~~~~~~~~~~~-9.002 \\
&&~~~~~~~~~~~~1096.00&20.90 & $^{94}$Mo($\alpha$,3n) & ~~~~~~~~~~~~-26.748\\
&&~~~~~~~~~~~~626.83 & 17.80\\
&&~~~~~~~~~~~~1178.70 & 05.10\\
&&~~~~~~~~~~~~806.28 & 4.04\\ 
\vspace{1mm}
$^{97}$Ru &~~~~~~~~~~~~2.791 d &~~~~~~~~~~~~215.70 & 85.62 & $^{94}$Mo($\alpha$,n) & ~~~~~~~~~~~~-7.940\\
&&~~~~~~~~~~~~324.49&10.79 & $^{95}$Mo($\alpha$,2n) & ~~~~~~~~~~~~ -15.309\\
&&&& $^{96}$Mo($\alpha$,3n) &~~~~~~~~~~~~-24.463\\
&&&&$^{97}$Mo($\alpha$,4n) &~~~~~~~~~~~~-31.284\\

\vspace{1mm}
$^{103}$Ru &~~~~~~~~~~~~ 39.26 d & ~~~~~~~~~~~~497.20 & 91.00 & $^{100}$Mo($\alpha$,n) & ~~~~~~~~~~~~-4.572\\
&&~~~~~~~~~~~~610.33 & 5.76& $^{103}$Tc decay (100 \%)\\
\vspace{1mm}
$^{94g}$Tc &~~~~~~~~~~~~ 293 min & ~~~~~~~~~~~~702.7 & 99.6 & $^{92}$Mo($\alpha$,p+n) & ~~~~~~~~~~~~-15.586\\
&&~~~~~~~~~~~~849.7&95.7 & $^{94}$Mo($\alpha$,3n) & ~~~~~~~~~~~~-33.334\\
&&~~~~~~~~~~~~871.1&99.9 & $^{94m}$Tc decay ($<0.1\%$)\\
&&~~~~~~~~~~~~916.1&7.6\\
\vspace{1mm}
$^{95g}$Tc &~~~~~~~~~~~~ 20.0 h & ~~~~~~~~~~~~765.78 & 93.80 & $^{92}$Mo($\alpha$,p) & ~~~~~~~~~~~~-5.652\\
&&~~~~~~~~~~~~1073.71&3.74 & $^{94}$Mo($\alpha$,t) & ~~~~~~~~~~~~-14.918\\
&&~~~~~~~~~~~~947.67&1.95 & $^{94}$Mo($\alpha$,2n+p) & ~~~~~~~~~~~~  -23.399\\
&&&& $^{95}$Mo($\alpha$,3n+p) & ~~~~~~~~~~~~ -30.769\\
&&&&$^{95m}$Tc decay (3.88 \%)\\
&&&&$^{95}$Ru decay (97 \%)\\
\vspace{1mm}
$^{96g}$Tc &~~~~~~~~~~~~ 4.28 d & ~~~~~~~~~~~~778.2 & 99.76 & $^{94}$Mo($\alpha$,p+n) & ~~~~~~~~~~~~-15.528\\
&&~~~~~~~~~~~~812.5&82 & $^{95}$Mo($\alpha$,p+2n) & ~~~~~~~~~~~~-22.897\\
&&~~~~~~~~~~~~849.9 & 98 & $^{96}$Mo($\alpha$,p+3n) & ~~~~~~~~~~~~-32.051\\
&&~~~~~~~~~~~~1126.85&15.2 & $^{96m}$Tc decay (98\%) \\

\hline
\hline
\end{tabular}
\end{center}
\end{table*}

\subsection{$^{103}$Ru (T$_{1/2}$ = 39.26 d)}
The measured excitation function as a function of the incident alpha energy for the nuclear reaction $^{100}$Mo($\alpha$,n)$^{103}$Ru with available experimental data from the EXFOR and the TALYS nuclear code theoretical calculation is shown in Fig. 4.
The radionuclide $^{103}$Ru is produced through the nuclear reaction $^{100}$Mo($\alpha$,n)$^{103}$Ru and by the decay of $^{103}$Tc which has short half life of 54.2 sec. The  reaction cross-sections for the nuclear reaction $^{100}$Mo($\alpha$,n)$^{103}$Ru were obtained from a $\gamma$-ray of 497.08 keV energy and intensity I$_\gamma$ = 91 $\%$, which decay from $^{103}$Ru radionuclide. The counting of this $\gamma$-ray was done after cooling time of about 2 days. It is clear from Fig. 5 that present experimental data  for this reaction are found to be in good agreement with the reaction data reported by R. A. Esterlund+ et al.(1965), \cite{41}, but our results slightly lower than available experimental data reported by H. P. Graf+ et al.(1974), \cite{42}. The theoretical results (shown in the magenta colour by solid line) made using ldmodel-3 are best in line with the present experimental data and follow the trend of experimental results of this reaction. The measured reaction cross sections along with their uncertainties and covariance metrics for the reaction $^{100}$Mo($\alpha$,n)$^{103}$Ru are presented in Table IV. 

\begin{figure}[b]
\begin{center}
\includegraphics[width=8.0 cm]{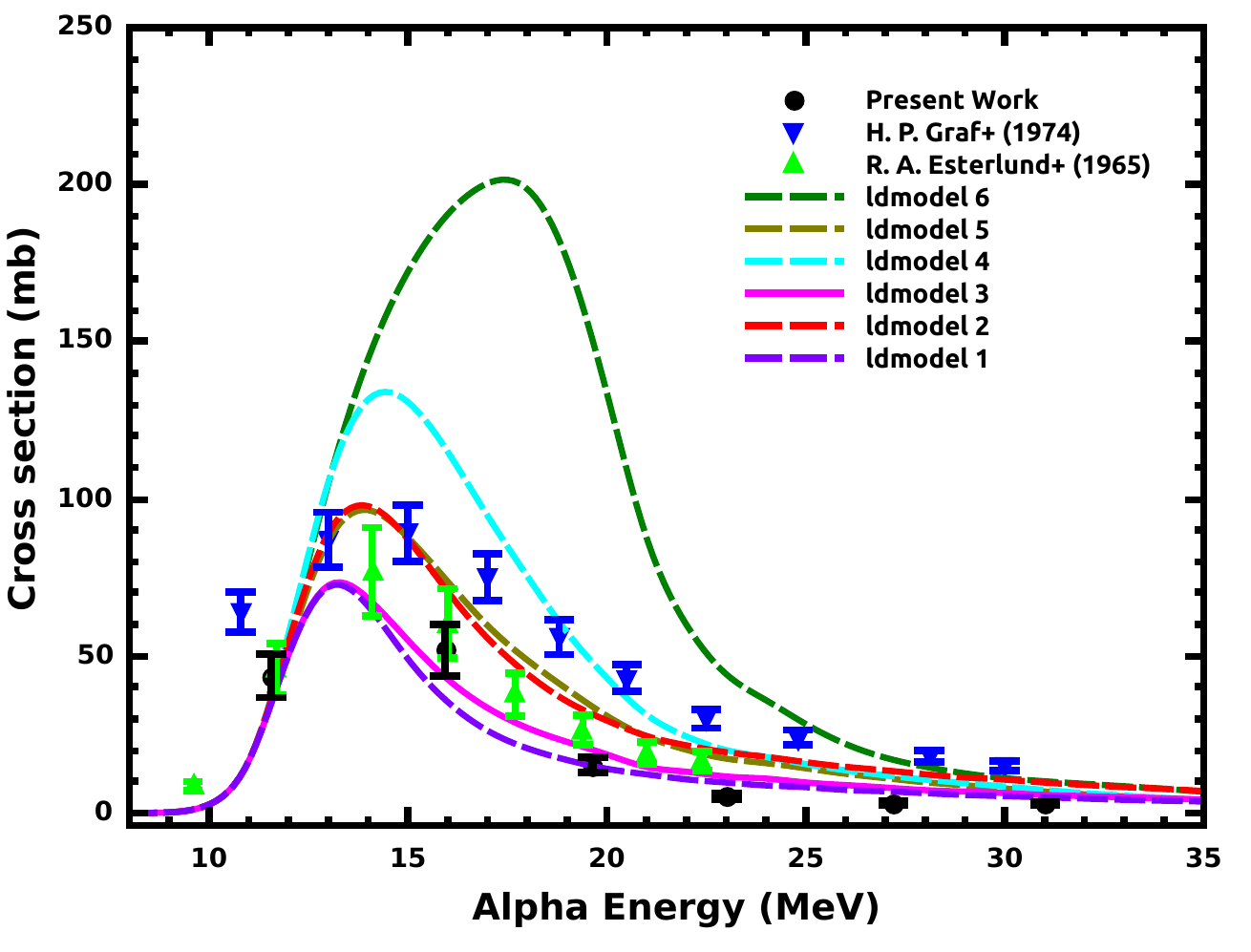}
\caption{The excitation function of reaction $^{100}$Mo($\alpha$,n)$^{103}$Ru along with available experimental data from EXFOR and theortical results from TALYS nuclear code.}
\end{center}
\end{figure}
\begin{figure}[b]
\begin{center}
\includegraphics[width=8.0 cm]{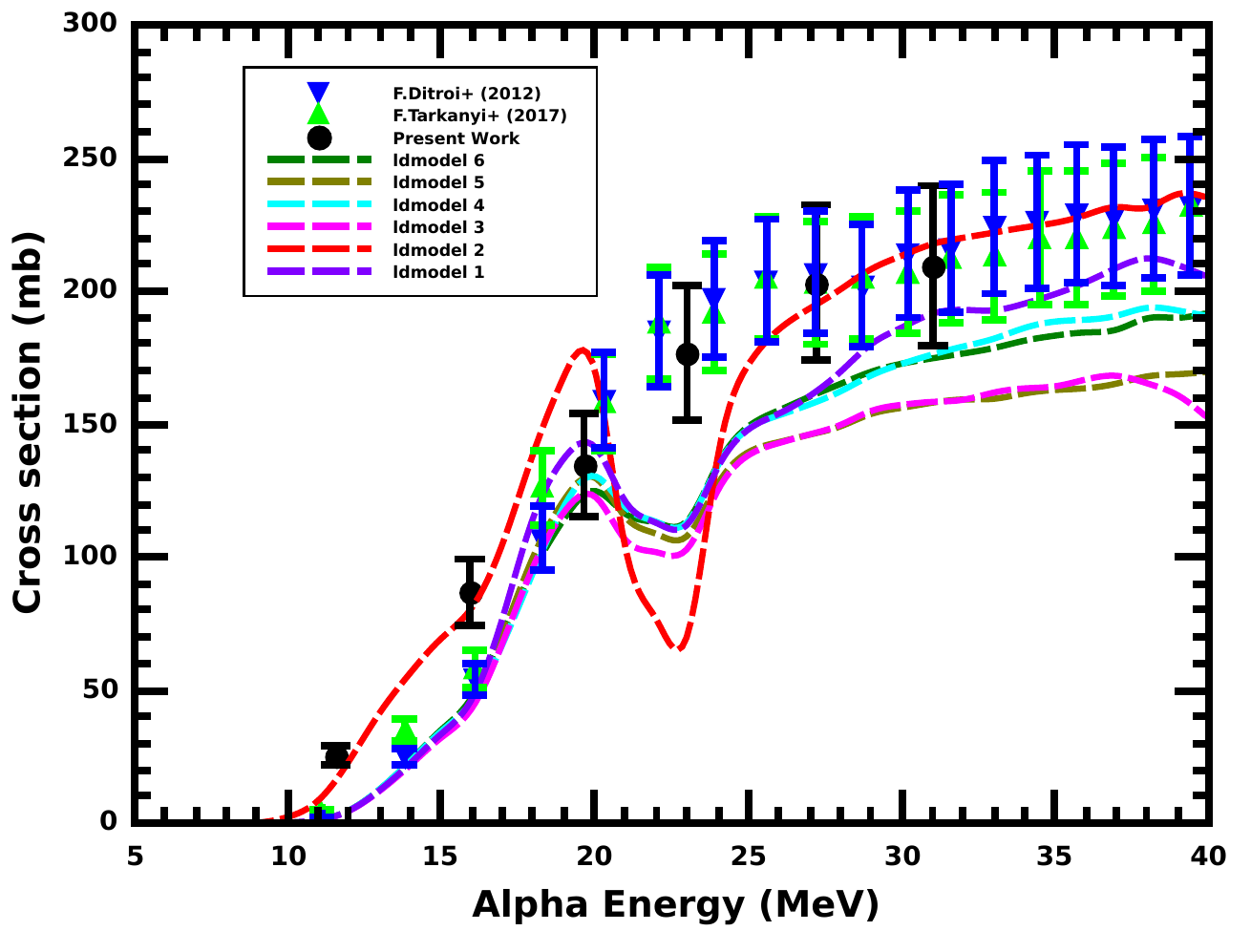}
\caption{The excitation function of reaction $^{nat}$Mo($\alpha$,x)$^{97}$Ru along with available experimental data from EXFOR and theortical results from TALYS nuclear code.}
\end{center}
\end{figure}

\begin{table*}[t]
\begin{center}
{\ TABLE IV. The measured reaction cross sections of the reaction $^{100}$Mo($\alpha$,n)$^{103}$Ru  with their uncertainties and covariance matrix. \\
\vspace{2mm}

\begin{tabular}{ccccccccccc}
\hline
\hline
E$_{\alpha}$ (MeV)&~~~~~~~~~~~~~~~~~~~Cross section (mb)&\multicolumn{4}{c}~~~~~~~~~~~~~~~{covariance matrix}\\
&~~~~~~~~~~~~~~~~~~~($\sigma$ $\pm$ $\bigtriangleup\sigma$)&&\\
\hline
\vspace{1mm}
11.56 & ~~~~~~~~~~~~~~~~~~~43.704 $\pm$ 6.941 & ~~~~~~~~~~~~~~~~~~~48.188 \\
\vspace{1mm}
15.93 &~~~~~~~~~~~~~~~~~~~51.831 $\pm$ 8.207 &~~~~~~~~~~~~~~~~~~~56.765 & ~67.364  &\\
\vspace{1mm}
19.65 & ~~~~~~~~~~~~~~~~~~~15.297 $\pm$ 2.448 & ~~~~~~~~~~~~~~~~~~~16.751 & ~19.869 & ~5.992 &\\
\vspace{1mm}
23.01 & ~~~~~~~~~~~~~~~~~~~5.353 $\pm$ 0.871 & ~~~~~~~~~~~~~~~~~~~5.861 & ~6.952 & ~2.051 & ~0.759\\
\vspace{1mm}
27.22 & ~~~~~~~~~~~~~~~~~~~3.171 $\pm$ 0.509 & ~~~~~~~~~~~~~~~~~~~3.473 & ~4.119 & ~1.215 & ~0.425 & ~0.259\\
31.01 & ~~~~~~~~~~~~~~~~~~~3.043 $\pm$ 0.492 & ~~~~~~~~~~~~~~~~~~~3.333 & ~3.953 & ~1.166 & ~0.408 & ~0.241 & ~0.242\\
\hline
\hline
\end{tabular}}
\end{center}
\end{table*}
\begin{table*}[t]
\begin{center}
{\ TABLE V. The measured reaction cross sections of the reaction $^{nat}$Mo($\alpha$,x)$^{97}$Ru  with their uncertainties and covariance matrix. \\
\vspace{2mm}

\begin{tabular}{ccccccccccc}
\hline
\hline
E$_{\alpha}$ (MeV)&~~~~~~~~~~~~~~~~~~~Cross section (mb)&\multicolumn{4}{c}~~~~~~~~~~~~~~~{covariance matrix}\\
&~~~~~~~~~~~~~~~~~~~($\sigma$ $\pm$ $\bigtriangleup\sigma$)&&\\
\hline
\vspace{1mm}
11.56 & ~~~~~~~~~~~~~~~~~~~25.422 $\pm$ 3.640 & ~~~~~~~~~~~~~~~~~~~13.256 \\
\vspace{1mm}
15.93 & ~~~~~~~~~~~~~~~~~~~86.794 $\pm$ 12.430 & ~~~~~~~~~~~~~~~~~~~45.253 & ~154.511  &\\
\vspace{1mm}
19.65 & ~~~~~~~~~~~~~~~~~~~134.539 $\pm$ 19.269 & ~~~~~~~~~~~~~~~~~~~70.147 & ~239.490 & ~371.317 &\\
\vspace{1mm}
23.01 & ~~~~~~~~~~~~~~~~~~~176.741 $\pm$ 25.314 & ~~~~~~~~~~~~~~~~~~~92.151 & ~314.615 & ~487.683 & ~640.808\\
\vspace{1mm}
27.22 & ~~~~~~~~~~~~~~~~~~~203.241 $\pm$ 29.108 & ~~~~~~~~~~~~~~~~~~~105.968 & ~361.787 & ~560.804 & ~736.719 & ~847.294\\
31.01 & ~~~~~~~~~~~~~~~~~~~209.541 $\pm$ 30.010 & ~~~~~~~~~~~~~~~~~~~109.252 & ~373.000 & ~578.186 & ~759.553 & ~873.438 & ~900.654\\
\hline
\hline
\end{tabular}}
\end{center}
\end{table*}

\subsection{$^{97}$Ru(T$_{1/2}$ = 2.83 d)}
The measured excitation function as a function of the incident alpha energy for the nuclear reaction $^{nat}$Mo($\alpha$,x)$^{97}$Ru with available experimental data from the EXFOR and the TALYS nuclear code theoretical calculation is shown in Fig. 6. The reaction cross-sections for the nuclear reaction $^{nat}$Mo($\alpha$,x)$^{97}$Ru were obtained from a $\gamma$-ray of 215.70 keV energy and intensity I$_\gamma$ = 85.62 $\%$, which decay from $^{97}$Ru radionuclide. The counting of this $\gamma$-ray was done after cooling time of about
2 days. From Fig. 6 it is clear that present experimental data  for this reaction are found to be in good agreement with the reaction data reported by  F.Tarkanyi+ et al (2017), and  F.Ditroi+ et al (2012) \cite{43, 44}. The theoretical results (shown in the red solid line) made using ldmodel-2 are best in line with the present experimental data and follow the trend of experimental results of this reaction but theortical results are lower than experimental data in the energy range 21-24 MeV. The measured reaction cross sections along with their uncertainties and covariance metrics for the reaction $^{nat}$Mo($\alpha$,x)$^{97}$Ru are presented in Table V.


\begin{table*}[t]
\begin{center}
{\ TABLE VI. The measured reaction cross sections of the reaction $^{nat}$Mo($\alpha$,x)$^{95}$Ru  with their uncertainties and covariance matrix. \\
\vspace{2mm}

\begin{tabular}{ccccccccccc}
\hline
\hline
E$_{\alpha}$ (MeV)&~~~~~~~~~~~~~~~~~~~Cross section (mb)&\multicolumn{4}{c}~~~~~~~~~~~~~~~~{covariance matrix}\\
&~~~~~~~~~~~~~~~~~~~($\sigma$ $\pm$ $\bigtriangleup\sigma$)&&\\
\hline
\vspace{1mm}
11.56 & ~~~~~~~~~~~~~~~~~~~11.609 $\pm$ 1.264 & ~~~~~~~~~~~~~~~~~~~1.599 \\
\vspace{1mm}
15.93 & ~~~~~~~~~~~~~~~~~~~52.467 $\pm$ 5.158 & ~~~~~~~~~~~~~~~~~~~5.869 & ~26.609  &\\
\vspace{1mm}
19.65 & ~~~~~~~~~~~~~~~~~~~87.196 $\pm$ 8.855 & ~~~~~~~~~~~~~~~~~~~9.753 & ~44.083 & ~78.411 &\\
\vspace{1mm}
23.01 & ~~~~~~~~~~~~~~~~~~~ 39.708 $\pm$ 4.056 & ~~~~~~~~~~~~~~~~~~~4.441 & ~20.074 & ~33.362 & ~16.451\\
\vspace{1mm}
27.22 & ~~~~~~~~~~~~~~~~~~~ 12.160 $\pm$ 1.276 & ~~~~~~~~~~~~~~~~~~~1.360 & ~6.147 & ~10.216 & ~4.652 & ~1.628\\
31.01 & ~~~~~~~~~~~~~~~~~~~ 4.669 $\pm$ 0.521 & ~~~~~~~~~~~~~~~~~~~0.522 & ~2.360 & ~3.923 & ~1.786 & ~0.547 & ~0.272\\
\hline
\hline
\end{tabular}}
\end{center}
\end{table*}
\begin{table*}[t]
\begin{center}
{\ TABLE VII. The measured reaction cross sections of the reaction $^{nat}$Mo($\alpha$,x)$^{96g}$Tc with their uncertainties and covariance matrix. \\
\vspace{2mm}

\begin{tabular}{cccccccccccc}
\hline
\hline
E$_{\alpha}$ (MeV)&~~~~~~~~~~~~~~~~~~~Cross section (mb)&\multicolumn{4}{c}~~~~~~~{covariance matrix}~~~~~~~~~~\\
&~~~~~~~~~~~~~~~~~~~($\sigma$ $\pm$ $\bigtriangleup\sigma$)&&\\
\hline
\vspace{1mm}
19.65 & ~~~~~~~~~~~~~~~~~~~0.256 $\pm$ 0.071 & ~~~~~~~~~~~~~~~~~~~0.005\\
\vspace{1mm}
23.01 & ~~~~~~~~~~~~~~~~~~~1.345 $\pm$ 0.132 & ~~~~~~~~~~~~~~~~~~~0.008 & ~~~0.017\\
\vspace{1mm}
27.22 & ~~~~~~~~~~~~~~~~~~~21.119 $\pm$ 1.975 & ~~~~~~~~~~~~~~~~~~~0.131 & ~~~0.261 & ~~~3.903\\
31.01 & ~~~~~~~~~~~~~~~~~~~36.412 $\pm$ 3.391 & ~~~~~~~~~~~~~~~~~~~0.226 & ~~~0.450 & ~~~6.660 & ~~~11.504\\
\hline
\hline
\end{tabular}}
\end{center}
\end{table*}

\subsection{$^{95}$Ru(T$_{1/2}$ = 1.643 h)}
In this work, Fig. 7 shows the measured excitation function as a function of the incident alpha energy in the energy range 11– 32 MeV of the nuclear  reaction $^{nat}$Mo($\alpha$,x)$^{95}$Ru with available experimental data from the EXFOR and the TALYS nuclear code theoretical calculation.
The reaction cross-sections for the nuclear reaction $^{nat}$Mo($\alpha$,x)$^{95}$Ru were obtained from a $\gamma$-ray of 336.40 keV energy and intensity I$_\gamma$ = 69.9 $\%$, which decay from $^{95}$Ru radionuclide. The counting of this $\gamma$-ray was done after a short cooling time. It is clear from Fig. 7 that for $^{nat}$Mo($\alpha$,x)$^{95}$Ru reaction the maximum value of cross-sections at incident alpha energy is found at about 18 MeV. The present experimental data for this reaction follow the trend of the experimental data reported by F.Tarkanyi+ et al.(2017), and  F.Ditroi+ et al(2012) \cite{43, 44}, but our results are slightly less than the available experimental data in the incident energy range 17–25 MeV. The theoretical results (shown in the red solid line) made using ldmodel-2 are best in line with the present experimental data and follow the trend of experimental results of this reaction. The measured reaction cross sections along with their uncertainties and covariance metrics for the reaction $^{nat}$Mo($\alpha$,x)$^{95}$Ru are presented in Table VI.\\ 


\begin{table*}[t]
\begin{center}
{\ TABLE VIII. The measured reaction cross sections of the reaction $^{nat}$Mo($\alpha$,x)$^{95g}$Tc  with their uncertainties and covariance matrix. \\
\vspace{2mm}

\begin{tabular}{ccccccccccc}
\hline
\hline
E$_{\alpha}$ (MeV)&~~~~~~~~~~~~~~~~~~~Cross section (mb)&\multicolumn{4}{c}~~~~~~~~~~~~~~~{covariance matrix}\\
&~~~~~~~~~~~~~~~~~~~($\sigma$ $\pm$ $\bigtriangleup\sigma$)&&\\
\hline
\vspace{1mm}
11.56 & ~~~~~~~~~~~~~~~~~~~6.476 $\pm$ 0.790 & ~~~~~~~~~~~~~~~~~~~0.624\\
\vspace{1mm}
15.93 & ~~~~~~~~~~~~~~~~~~~55.844 $\pm$ 6.675 & ~~~~~~~~~~~~~~~~~~~5.145 & ~44.559  &\\
\vspace{1mm}
19.65 & ~~~~~~~~~~~~~~~~~~~100.185 $\pm$ 12.010 & ~~~~~~~~~~~~~~~~~~~9.241 & ~79.821 & ~144.246 &\\
\vspace{1mm}
23.01 & ~~~~~~~~~~~~~~~~~~~ 72.122 $\pm$ 8.660 & ~~~~~~~~~~~~~~~~~~~6.662 & ~57.547 & ~103.398 & ~74.995\\
\vspace{1mm}
27.22 & ~~~~~~~~~~~~~~~~~~~ 25.325 $\pm$ 3.043 & ~~~~~~~~~~~~~~~~~~~2.342 & ~20.237 & ~36.365 & ~26.225 & ~9.260\\
31.01 & ~~~~~~~~~~~~~~~~~~~ 11.026 $\pm$ 1.380 & ~~~~~~~~~~~~~~~~~~~1.021 & ~8.823 & ~15.856 & ~11.436 & ~4.023 & ~1.905\\
\hline
\hline
\end{tabular}}
\end{center}
\end{table*}

\subsection{$^{96g}$Tc, (T$_{1/2}$ = 4.28 d )}
The measured excitation function as a function of the incident alpha energy for the nuclear reaction $^{nat}$Mo($\alpha$,x)$^{96g}$Tc with available experimental data from the EXFOR and the TALYS nuclear code theoretical calculation is shown in Fig. 7. The radionuclide $^{96g}$Tc is produced through the nuclear reactions $^{nat}$Mo($\alpha$,x) and by the internal decay of $^{96m}$Tc which has half life of 52 min. The reaction cross-sections for the nuclear reaction $^{nat}$Mo($\alpha$,x)$^{96g}$Tc were obtained from a $\gamma$-ray of 313.23 keV energy and intensity I$_\gamma$ = 2.43 $\%$.

\begin{figure}[b]
\begin{center}
\includegraphics[width=8.0 cm]{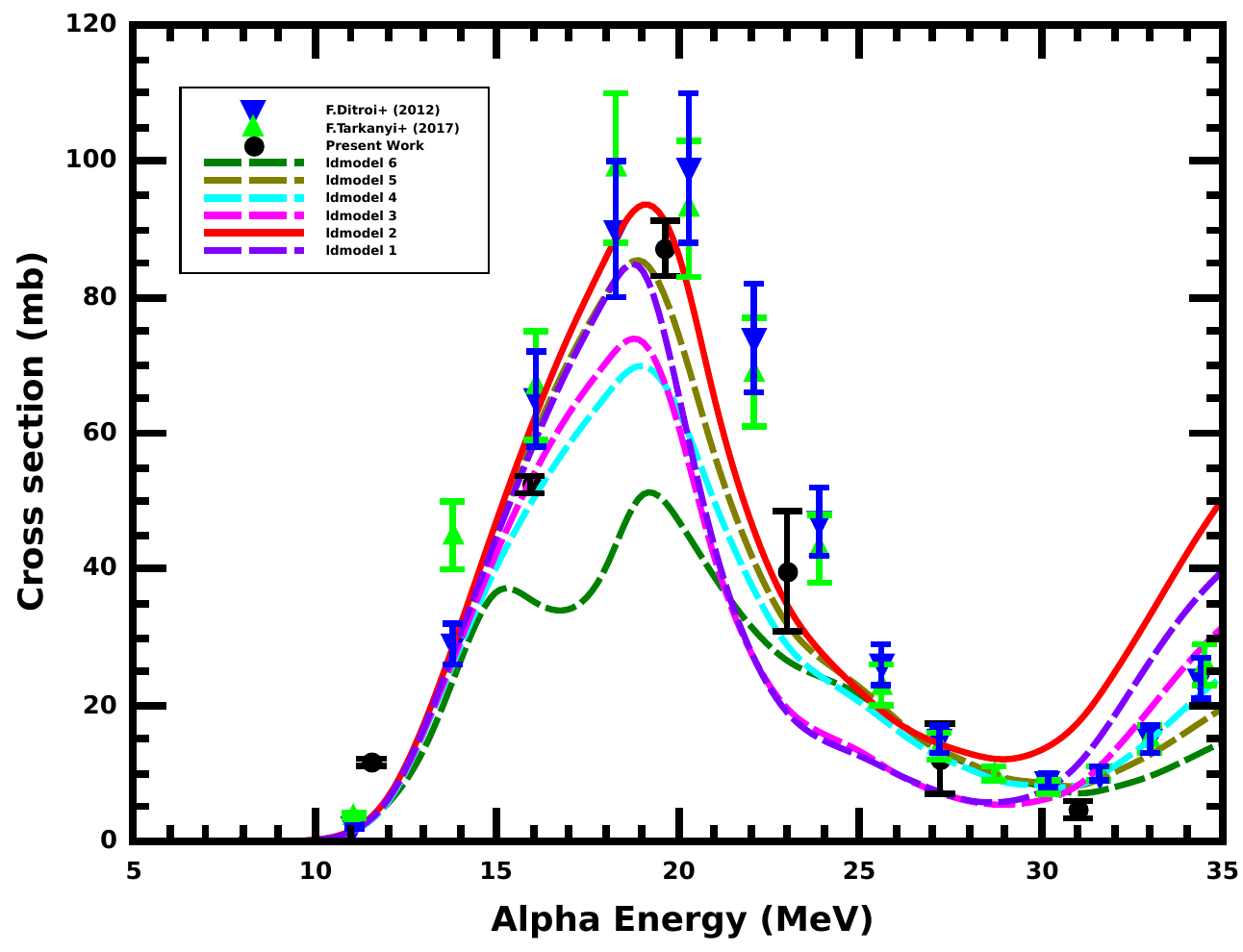}
\caption{The excitation function of reaction $^{nat}$Mo($\alpha$,x)$^{95}$Ru along with available experimental data from EXFOR and theortical results from TALYS nuclear code.}
\end{center}
\end{figure}

The counting of this $\gamma$-ray was done after cooling time of about 2 days, leading to complete decay of the parent. There is no experimental data available in the EXFOR data library for the nuclear reaction $^{nat}$Mo($\alpha$,x)$^{96g}$Tc. It is clear from Fig. 7 that the present experimental data of this nuclear reaction are found in good agreement with the theoretical results made using ldmodel-2. The measured reaction cross sections along with their uncertainties and covariance metrics for the nuclear reaction $^{nat}$Mo($\alpha$,x)$^{96g}$Tc are presented in Table VII. 
\begin{figure}[b]
\begin{center}
\includegraphics[width=8.0 cm]{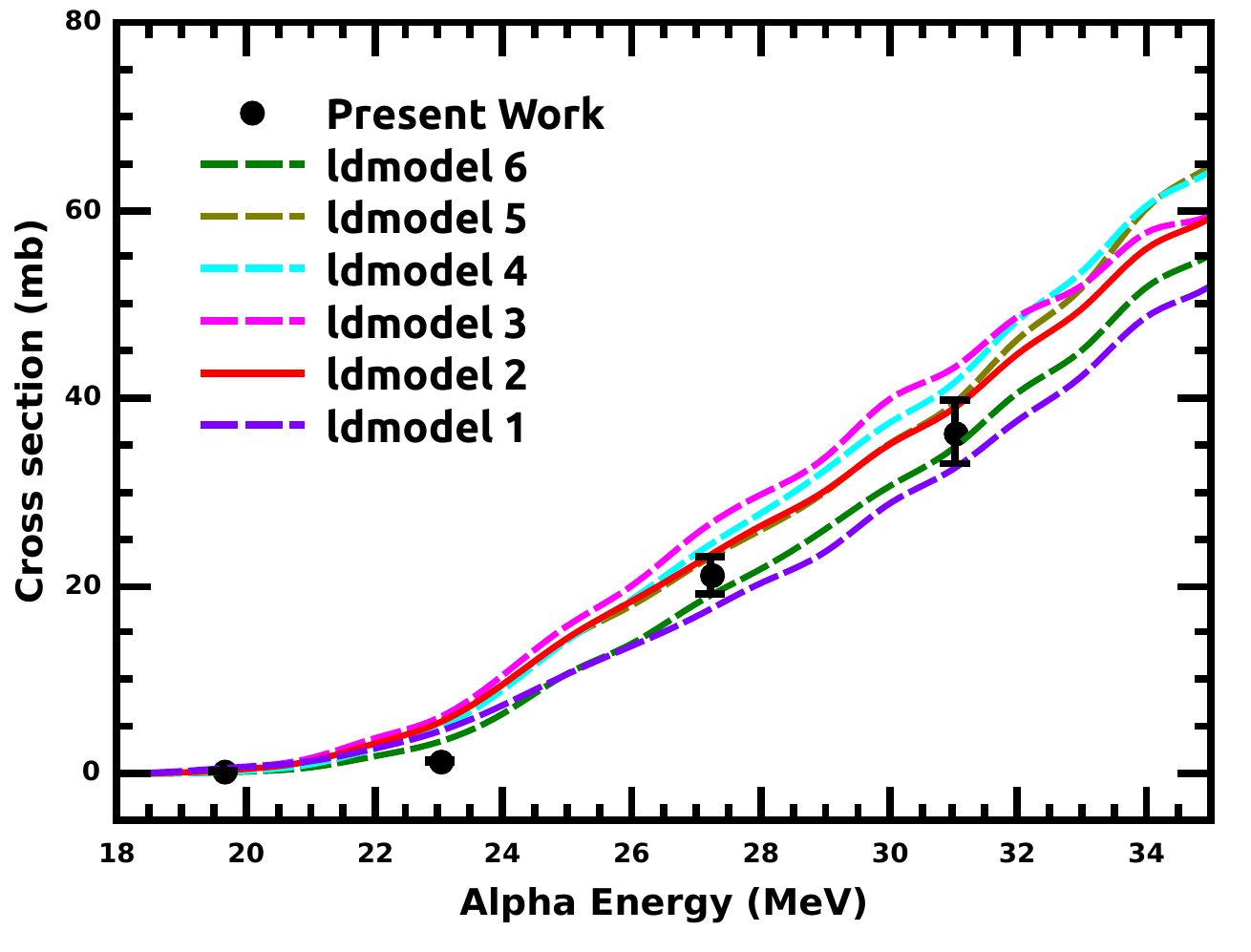}
\caption{The excitation function of reaction $^{nat}$Mo($\alpha$,x)$^{96g}$Tc along with available experimental data from EXFOR and theortical results from TALYS nuclear code.}
\end{center}
\end{figure}
\begin{figure}[b]
\begin{center}
\includegraphics[width=8.0 cm]{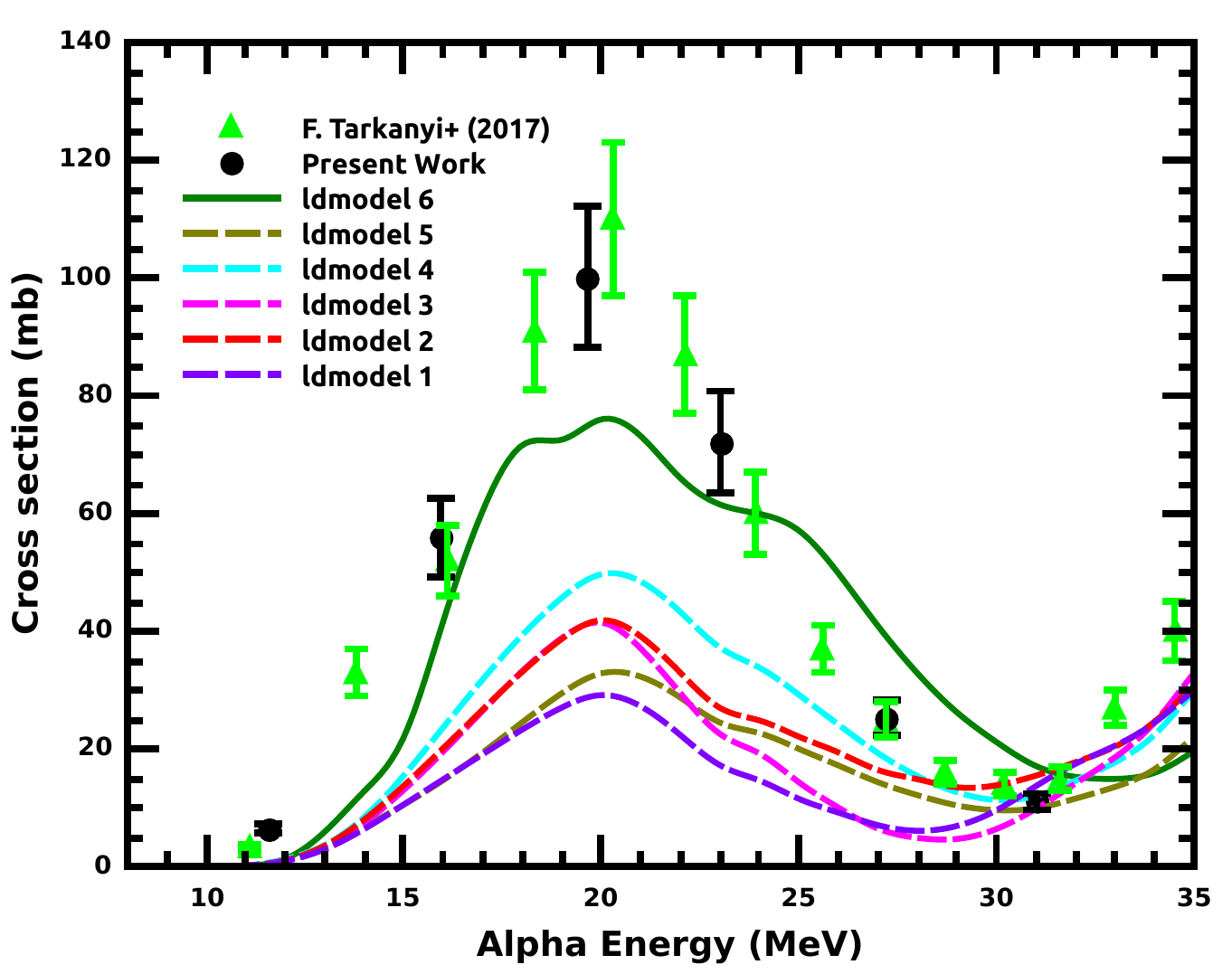}
\caption{The excitation function of reaction $^{nat}$Mo($\alpha$,x)$^{95g}$Tc along with available experimental data from EXFOR and theortical results from TALYS nuclear code.}
\end{center}
\end{figure}


\begin{table*}[t]
\begin{center}
{\ TABLE IX. The measured reaction cross sections of the reaction $^{nat}$Mo($\alpha$,x)$^{94g}$Tc  with their uncertainties and covariance matrix. \\
\vspace{2mm}

\begin{tabular}{ccccccccccc}
\hline
\hline
E$_{\alpha}$ (MeV)&~~~~~~~~~~~~~~~~~~~Cross section (mb)&\multicolumn{4}{c}~~~~~~~~~~~~~~~{covariance matrix}~~~~~~~~~~~~~~~\\
&~~~~~~~~~~~~~~~~~~~($\sigma$ $\pm$ $\bigtriangleup\sigma$)&&\\
\hline
\vspace{1mm}
19.65 & ~~~~~~~~~~~~~~~~~~~1.096 $\pm$ 0.254 & ~~~~~~~~~~~~~~~~~~~0.064\\
\vspace{1mm}
23.01 & ~~~~~~~~~~~~~~~~~~~43.347 $\pm$ 9.550 & ~~~~~~~~~~~~~~~~~~~2.298 & ~~~91.210\\
\vspace{1mm}
27.22 & ~~~~~~~~~~~~~~~~~~~76.425 $\pm$ 16.810 & ~~~~~~~~~~~~~~~~~~~4.051 & ~~~160.168 & ~~~282.606\\
31.01 & ~~~~~~~~~~~~~~~~~~~97.846 $\pm$ 21.518 & ~~~~~~~~~~~~~~~~~~~5.186 & ~~~205.043 & ~~~361.490 & ~~~463.061\\
\hline
\hline
\end{tabular}}
\end{center}
\end{table*}

\subsection{$^{95g}$Tc(cum), (T$_{1/2}$ = 20 h)}
The measured excitation function as a function of the incident alpha energy in the energy range 11– 32 MeV of the reaction $^{nat}$Mo($\alpha$,x)$^{95g}$Tc with available experimental data from the EXFOR and the TALYS nuclear code theoretical calculation is shown in Fig. 8. The reaction cross-sections for the nuclear reaction $^{nat}$Mo($\alpha$,x)$^{95g}$Tc were obtained from a $\gamma$-ray of 1073.71 keV energy and intensity I$_\gamma$ = 3.74 $\%$. There is small contribution of 3.83 \% isomeric transition ($^{95m}$Tc, T$_{1/2}$ = 61 d). To minimize the contribution of this isomeric transition we have used short cooling time. During the measurement, special attention was also paid to ensure that the contribution of the short-lived parent $^{95}$Ru (T$_{1/2}$ = 1.65 h) should be minimum. From Fig. 8, it is clear that present experimental data  for this reaction are found to be in good agreement with the reaction data reported by F. Tarkanyi et al \cite{43}. The present experimental data follows the trend of the theoretical results made using ldmodel-6, but the theoretical results are slightly lower than the experimental data in the energy range of about 11–23 MeV and slightly higher in the energy range of about 24–32 MeV. The measured reaction cross sections along with their uncertainties and covariance metrics for the nuclear reaction $^{nat}$Mo($\alpha$,x)$^{95g}$Tc are presented in Table VIII.

\begin{figure}[b]
\begin{center}
\includegraphics[width=8.8 cm]{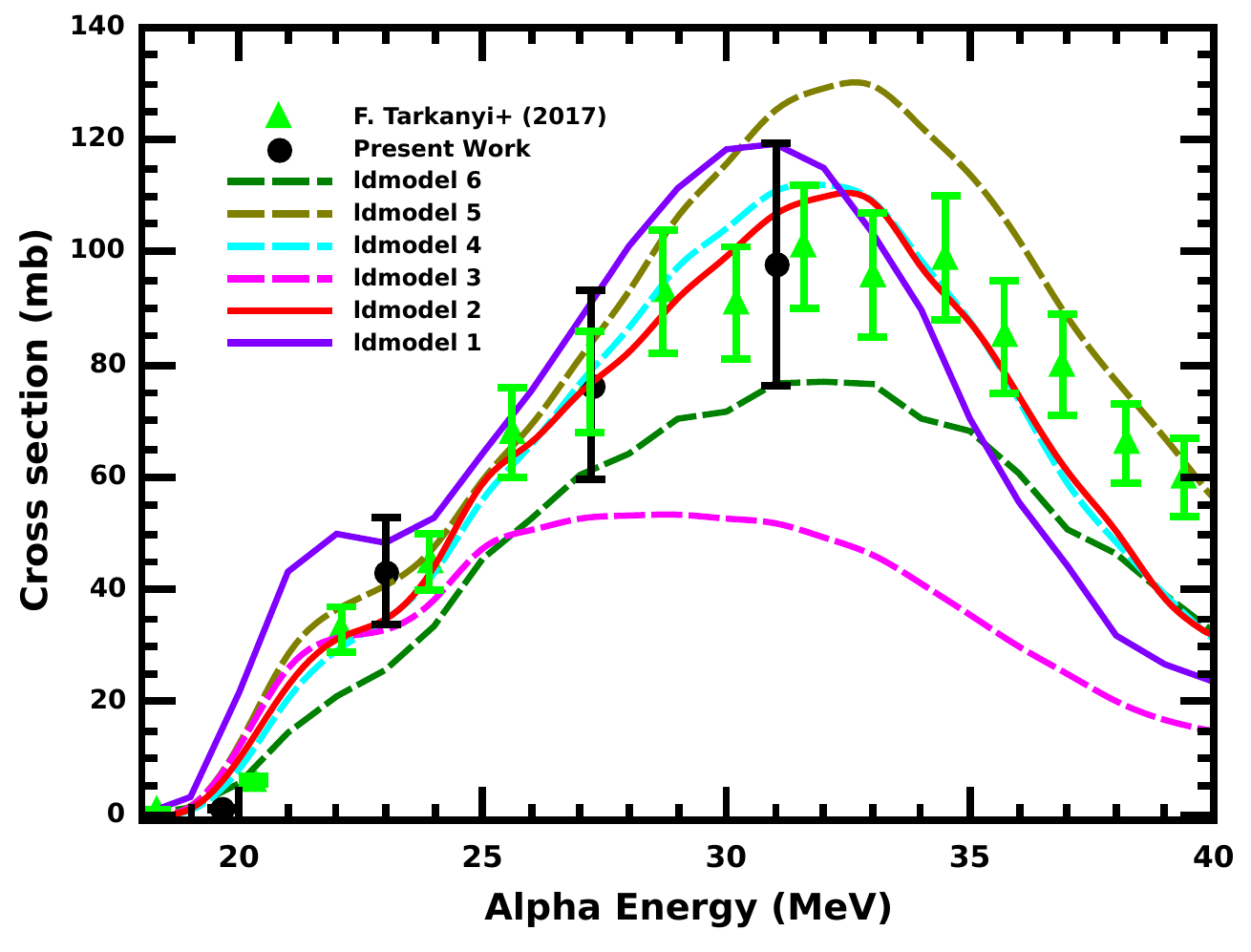}
\caption{The excitation function of reaction $^{nat}$Mo($\alpha$,x)$^{94g}$Tc along with available experimental data from EXFOR and theortical results from TALYS nuclear code.}
\end{center}
\end{figure}

\subsection{$^{94g}$Tc, (T$_{1/2}$ = 293 min)}
The measured excitation function as a function of the incident alpha energy for the nuclear reaction $^{nat}$Mo($\alpha$,x)$^{94g}$Tc with available experimental data from the EXFOR and the TALYS nuclear code theoretical calculation is shown in Fig. 9. The radionuclide $^{94g}$Tc is produced through the nuclear reactions $^{nat}$Mo($\alpha$,x) and by the decay of $^{94}$Ru which has half life of 51.8 min.The short-lived isomeric state ($^{94m}$Tc) has a very small contribution (less than 0.1\%) that can be neglected. The reaction cross-sections for the nuclear reaction $^{nat}$Mo($\alpha$,x)$^{94g}$Tc were obtained from a $\gamma$-ray of 916.1 keV energy (I$_\gamma$ = 7.6 $\%$) and The activity was measured of this $\gamma$-ray after the complete decay of the parent. From Fig. 9, it is clear that present experimental data  for this reaction are found to be in good agreement with the reaction data reported by F. Tarkanyi et al \cite{43}. The present experimental data are in good agreement with the theoretical results made using ldmodel-2 and ldmodel-4. The measured reaction cross sections along with their uncertainties and covariance metrics for the reaction $^{nat}$Mo($\alpha$,x)$^{94g}$Tc are presented in Table IX.
\section{Conclusion} 
In the present work, the reaction cross-sections of $^{nat}$Mo($\alpha$,n)$^{103}$Ru, $^{nat}$Mo($\alpha$,x)$^{97}$Ru, $^{nat}$Mo($\alpha$,x)$^{95}$Ru,  $^{nat}$Mo($\alpha$,x)$^{96g}$Tc, $^{nat}$Mo($\alpha$,x)$^{95g}$Tc and $^{nat}$Mo($\alpha$,x)$^{94g}$Tc nuclear reactions in the energy range about  11-32 MeV are determined using stack foil activation technique along with covariance analysis. The documentation of detailed uncertainty analysis for these nuclear reactions and their corresponding covariance matrix are presented for the first time. The reaction cross-sections obtained for the above nuclear reactions are found to be consistent with experimental data available in the EXFOR data library and theoretical prediction from the TALYS nuclear code. The ldmodel-2 gives the best theoretical results for the nuclear reactions $^{nat}$Mo($\alpha$,x)$^{97}$Ru, $^{nat}$Mo($\alpha$,x)$^{95}$Ru, $^{nat}$Mo($\alpha$,x)$^{96g}$Tc, $^{nat}$Mo($\alpha$,x)$^{94g}$Tc, the ldmodel-3 gives the best theortical results for the nuclear reaction $^{100}$Mo($\alpha$,n)$^{103}$Ru and the ldmodel-6 gives the best theoretical results for the nuclear reactions  $^{nat}$Mo($\alpha$,x)$^{95g}$Tc. 
The present data analysis of the measured reaction cross sections took into account the necessary improvements as a result of the coincidence summing factor and the geometric factor in efficiency. The present experimental results with detailed covariance analysis are important in astrophysics, nuclear medicine, and improving the nuclear reaction codes.
\begin{acknowledgments}
The author (Mahesh Choudhary) gratefully thanks to the Council of Scientific and Industrial Research (CSIR), Government of India, for financial support in the form of Senior Research Fellowships. (File No 09/013(882)/2019-EMR-1). Also, one of the authors (A. Kumar) would like to thank the SERB, DST, Government of India [Grant No. CRG/2019/000360], and Institutions of Eminence (IoE) BHU [Grant No. 6031].\\
We acknowledge the kind support provided by Prof. A. K. Tyagi, Director, Chemistry Group, BARC, Mumbai and Prof. Chandana Bhattacharya, Head, Experimental Nuclear Physics Division, VECC, Kolkata towards the successful execution of the experiment. We would also like to thank the Cyclotron (K-130) staff, VECC, Kolkata for providing us high quality of the beam during the experiment.

\end{acknowledgments}
\bibliography{basename of .bib file}

\end{document}